\documentclass[12pt]{iopart}

\usepackage{upgreek}
 \usepackage{graphicx}
\usepackage{xcolor}
\begin{document}

\title{Wide-Field Strain Imaging with Preferentially-Aligned Nitrogen-Vacancy Centers in Polycrystalline Diamond }

\author{Matthew E. Trusheim$^1$ and Dirk Englund$^1$}
\address{$^1$Dept. of Electrical Engineering and Computer Science, MIT, Cambridge, MA 02139}
\eads{mtrush@mit.edu}

\date{\today}

\begin{abstract}
We report on wide-field optically detected magnetic resonance imaging of nitrogen-vacancy centers (NVs) in type IIa polycrystalline diamond. These studies reveal a heterogeneous crystalline environment that produces a varied density of NV centers, including preferential orientation within some individual crystal grains, but preserves long spin coherence times. Using the native NVs as nanoscale sensors, we introduce a 3-dimensional strain imaging technique with high sensitivity ( $< 10^{-5}$ Hz$^{-1/2}$) and diffraction-limited resolution across a wide field of view. 


\end{abstract}


\maketitle

\section{Introduction}
Recent years have seen rapid advances in the development of quantum memories and sensors based on solid-state spin systems. At the forefront of these spin systems is the nitrogen vacancy center in diamond (NV), which has an electron spin triplet ground state with exceptionally long coherence time even at room temperature \cite{Bar-Gill2013}. Measuring these spins by optically detected magnetic resonance (ODMR) has enabled high performance sensing of electromagnetic fields \cite{maze_nanoscale_2008,Dolde2011}, temperature \cite{Kucsko2013a} and pressure\cite{Doherty2014}. The highest-sensitivity experiments have used single-crystal diamond grown through chemical vapor deposition. However, such samples remain expensive and small (on the scale of millimeters), which limits the adoption and scaling of NV sensing techniques. Polycrystalline diamond (PCD) presents an attractive alternative as it can be grown on the wafer scale and at lower cost, with the potential for long spin coherence times\cite{Balmer2009,Jahnke2012}. However, a major concern with PCD has been that the diverse crystal structure, including grain boundaries and varied growth regimes, could lead to inconsistent and poor NV properties. In this work, we employ wide-field ODMR spectroscopy \cite{LeSage2013,Steinert2013,DeVience2015a,Chen2013} to characterize the properties of hundreds of individually-resolved NVs across a field of view of $> 300$ $\upmu$m$^2$  and to use those centers as high-resolution nanoscale sensors of local crystal strain. These studies reveal that NVs in PCD are naturally preferentially aligned in some crystal grains, have spatially varying concentration, can be strongly strained near grain boundaries, and exhibit consistently long spin coherence times. Using these NVs, we introduce a method for wide-field strain imaging. We produce detailed, 3-dimensional strain maps of PCD structure with diffraction-limited resolution and sensitivity below $10^{-5}$ Hz$^{-1/2}$, outperforming traditional optical strain measurement techniques such as Raman imaging\cite{Kato2012109}. These studies show the application of NVs for high-resolution strain imaging, and demonstrate the viability and potential advantages of polycrystalline diamond for quantum sensing and information processing. 

The NV is an electronic spin-1 system consisting of a substitutional nitrogen atom adjacent to a vacancy in the diamond lattice \cite{Doherty2012}. The ground-state spin can be coherently manipulated by microwave fields, as well as initialized and detected through optical illumination because the m$_s = \pm1$ sublevels are dark states that emit reduced fluorescence. The ground state Hamiltonian describing the system is : 

\begin{equation} H = \frac{1}{\hbar^2}[(D_{gs}+\mathcal{E} _z)S_z^2-\mathcal{E} _x(S_x^2-S_y^2)\\+ \mathcal{E} _y(S_xS_y+S_yS_x)]  +\frac{g\mu_b}{\hbar}\mathbf{S}\cdot \mathbf{B}
\end{equation}
Where $D_{gs}$ = 2.87 GHz is the spin-spin interaction energy, $\mathbf{\mathcal{E} } = \mathbf{d_{gs}}\cdot(\mathbf{E}+\mathbf{\sigma})$ is the energy of interaction with external electric ($\mathbf{E})$ and strain ($\mathbf{\sigma}$) fields, $\mathbf{B}$ is the magnetic field at the location of the NV, $d_{gs}$ is the NV electric dipole moment, $g$ the electron g-factor, and $\mu_b$ the Bohr magneton. Strain or electric fields along the $\vec{z}$ direction defined by the NV axis (Figure 1a) correspond to shifts in lattice spacing that preserve the NV's trigonal symmetry and affect both $m_s=\pm1$ spin levels equally, while the levels are split in the presence of non-axial strains ($\mathcal{E}_x, \mathcal{E}_y$) that break this symmetry. For low transverse magnetic fields  $B_{\perp} = \sqrt{B_x^2+B_y^2} << D_{gs}$, the ground state transition frequencies are

\begin{equation}\hbar\omega_{\pm} = D_{gs}+\mathcal{E} _z \pm \sqrt{\mathcal{E} _{\perp}^2+(g\mu_bB_z)^2}\end{equation}

Where $\mathcal{E} _{\perp}=\sqrt{\mathcal{E} _x^2+\mathcal{E} _y^2} $ is the perpendicular effective electric field. This expression indicates two limiting regimes corresponding to the dominance of the $B_z$ or $\mathcal{E} _{\perp}$ terms. In the high-field regime, $B_z \gg \mathcal{E} _{\perp}$, the NV resonance frequency is not sensitive to changes in $\mathcal{E} _{\perp}$ in first order. A large magnetic bias field therefore allows probing of NV orientation based on magnetic field projection along the NV axis. In the low-field regime, $ B_z \ll\mathcal{E} _{\perp}$, the NV is sensitive to changes in $\mathcal{E} _{\perp}$ but not $\mathbf{B}$. Operating in this regime by eliminating the bias magnetic field enables probing of local strain through measurement of both $\mathcal{E}_z$ and $\mathcal{E}_\perp$. The change in energy per unit strain $ \mathbf{\mathcal{E}} (\mathbf{\sigma})$ is anisotropic and has been measured in separate cantilever-based experiments \cite{Teissier2014,Ovartchaiyapong2014}  as well as under isotropic pressure\cite{Doherty2014}. For this work, we assume a shift of $\mathcal{E}_\perp = 20.6$ GHz and $\mathcal{E}_z= 9.38$ GHz derived from the mean of the reported values.

\begin{figure*}
	\includegraphics[width=17cm]{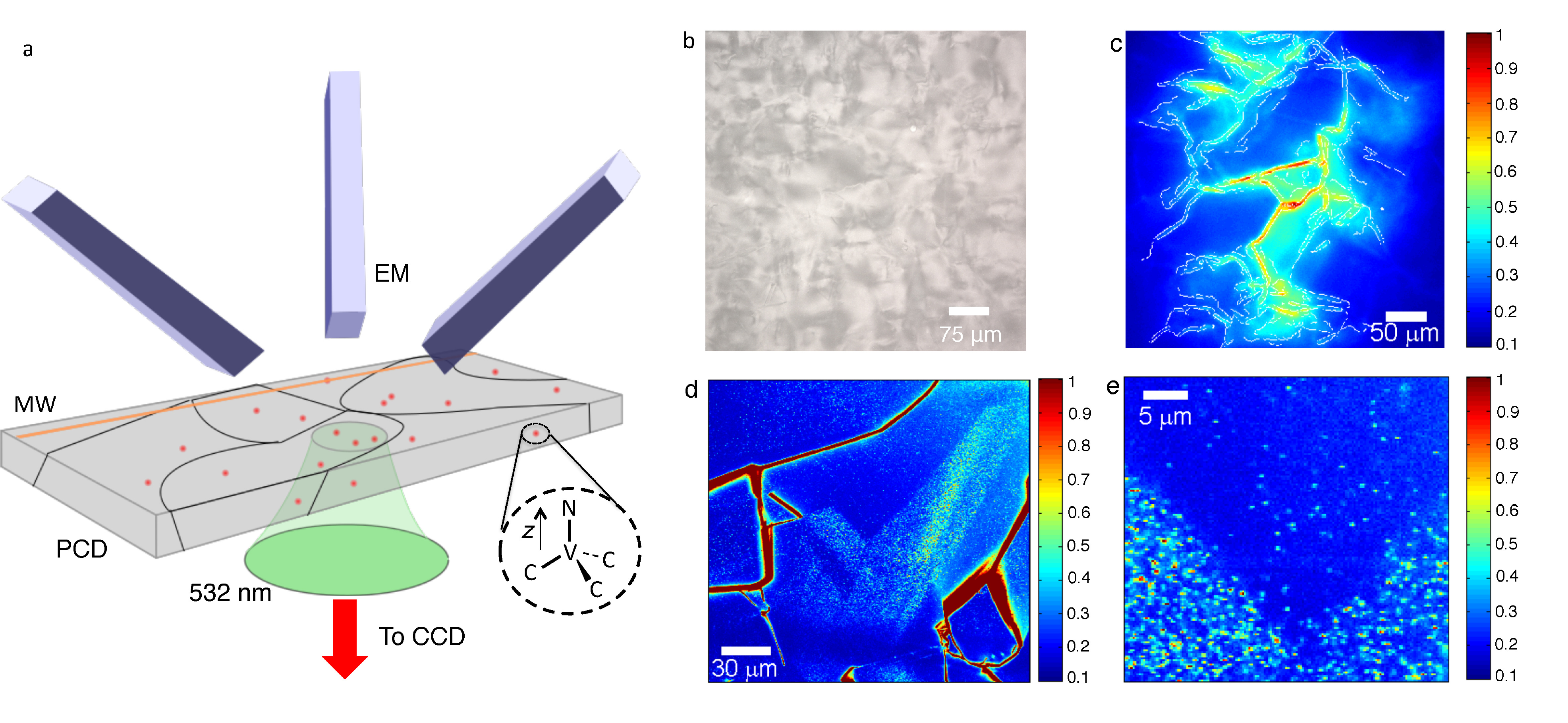}
	\caption{a) Experimental concept. The PCD sample is illuminated with 532 nm light, and the red fluorescence emitted by NVs embedded in the polycrystalline diamond is collected onto a CCD (not shown). NV electron spins are manipulated via microwave-frequency excitation applied from a wire on the diamond surface (MW), and the static magnetic field is controlled along three independent axes by electromagnets (EM). The NV symmetry axis defines the $z$ direction. b) Bright-field image of PCD. c) Wide-field image under 532 nm excitation. Bright fluorescence is seen on grain boundaries. The white lines serve as guides to the eye, outlining the crystal grains. d) Confocal scan taken at 100x magnification of a single grain. Within the grain, NV density varies depending on growth sector. e) Image of a density transition; each spot corresponds to a single NV. Color scale: relative intensity (arb. units). }
	\label{fig:spectrum}
\end{figure*}
\section{Wide-Field ODMR} 
To probe the properties of NVs in PCD, we performed wide-field ODMR measurements using a custom-built fluorescence microscope (Figure 1a). Optical illumination is provided by a 532 nm green laser modulated by a double-pass acousto-optic modulator, and resulting NV fluorescence is spectrally filtered (650 nm long-pass) and collected on an electron-multiplying CCD camera. Spin manipulation is accomplished by applying microwaves through a 15$\mu$m copper wire placed nearby to the area of interest, while the external magnetic field is controlled by three orthogonal current-controlled electromagnets in addition to a permanent rare-earth magnet. The measurements were performed at room temperature without external stabilization or control. 

We investigated a type-IIa polycrystalline diamond provided by Element6, grown by chemical vapor deposition with a nitrogen concentration $<$ 50 ppb. We first imaged the sample in fluorescence. Individual crystal grains vary in size from $~10-1000$ $\upmu$m$^2$ in area, and are easily identified in fluorescence imaging by their boundaries (Figure 1c,d). These grain boundaries fluoresce brightly across the visible band, likely due to a high density of optically-active lattice traps and amorphous carbon\cite{Jahnke2012}, and indicate the transition between two growth regimes. Individual NV centers within individual PCD grains are visible under 100 x magnification (Fig 1b,c). The measurements show an NV density that varies significantly within and between grains, from roughly $\sim 0.1$ NV/$\upmu$m$^2$ to densities approaching $ 1$ NV/$\upmu$m$^2$. This heterogeneity contrasts with the homogenous NV distribution in single-crystal CVD diamond.

\begin{figure}
	\centering\includegraphics[width=12cm]{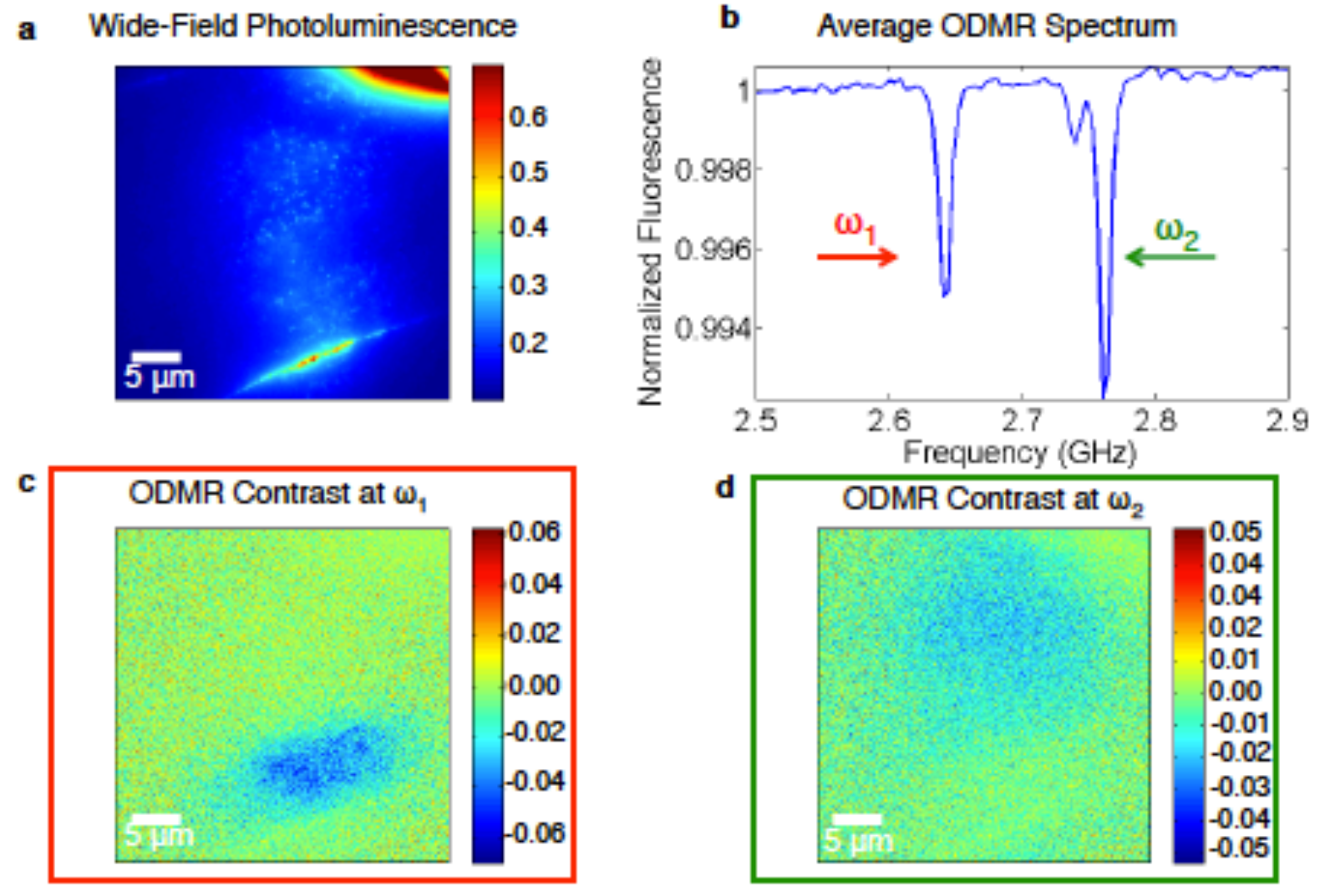}
	\caption{a) Wide-field fluorescence image of NVs in polycrystalline diamond. Each fluorescent spot corresponds to a single NV, and the background corresponds to many out-of-focus NVs. Bright grain boundaries are visible on the edges of the image. b) ODMR spectrum averaged over the field of view. Dimming corresponds to driving between the m$_s$ = 0 and m$_s$ = $-1$ spin sublevels of different geometric classes of NVs. c,d) ODMR contrast images corresponding to the resonance marked with the green (red) arrow in b. The geometric classes of NVs is spatially distinct.}
	\label{fig:spectrum}
\end{figure}

We then investigated NV properties in the high magnetic field regime. An external magnetic field of $\sim$ 100 G was applied and the ODMR spectrum of NVs within the field of view obtained under continuous-wave microwave and optical excitation (Figure 2a,b). The observed resonance frequencies correspond to specific geometric classes of NVs $\{i\}$, each with a defined angle to the external magnetic field resulting in a different resonance frequency $\hbar\omega_i = D_{gs}+g\mu_bB_{z(i)}$. Interestingly, only three distinct NV orientations appear in this region, rather than the four possible classes corresponding the four crystallographic (111) directions. In Figure 2c,d, we map the location of NV geometric classes $\{i=1,2\}$ by showing ODMR at frequencies $\omega_i$. This shows spatial separation of NV classes, likely by growth region as no intersecting grain boundary is observed, with the lower region only containing NVs at a single frequency. This effect persists independent of the direction of the applied permanent magnetic field (Supplemental Material), and across grains in the PCD. Preferential orientation of NVs through engineered growth has been theoretically predicted\cite{Miyazaki2014a,Karin2014a} and observed previously in single crystal diamond with controlled growth along \{110\}\cite{Pham2012a,Edmonds2012}, \{111\}\cite{fukui2014,Tahara2015} and \{113\}\cite{lesik2015preferential}. These results indicate that  preferential alignment occurs naturally in PCD, possibly due to predominant \{110\} and \{111\} grain textures.

\begin{figure*}
	\centering\includegraphics[width=17cm]{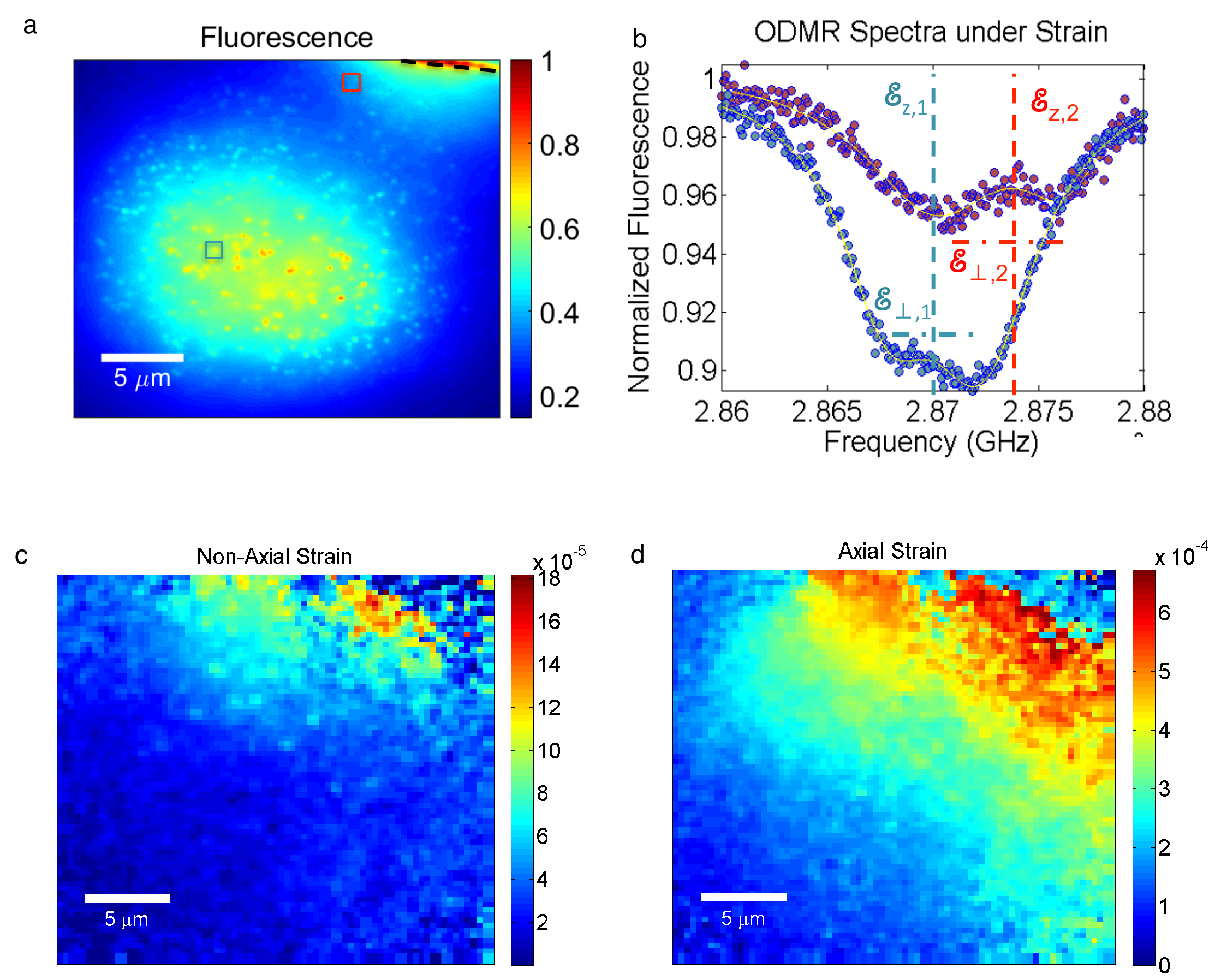}
	\caption{a) Wide-field fluorescence image of NVs in polycrystalline diamond. A grain boundary is visible in the upper-right corner indicated by the black dashed line. b) Low-field ODMR spectra corresponding to the locations indicated in a. The differences in $\mathcal{E}_z$ and $\mathcal{E}_\perp$ are indicated. c,d) Strain maps. Axial and non-axial strains both increase near the grain boundary. The data are binned in a 4x4 pixel ($\sim$320 nm) region and smoothed over single-pixel nearest neighbors.}
	\label{fig:spectrum}
\end{figure*}

\section{Strain Imaging}

We next turn to strain mapping of the PCD by wide-field continuous-wave ODMR spectroscopy. These measurements must be performed in the low magnetic field regime where $B_z \ll \mathcal{E}_\perp$; to achieve this, we canceled external magnetic fields using three-axis electromagnets. Figure 3b shows a representative ODMR spectrum. From double-Lorentzian fits to these low-field spectra, taken in parallel across the field of view with a total measurement time of 150 seconds, we extracted $(D_{gs}+\mathcal{E} _z)$ and  $\mathcal{E}_\perp$, following Equation 2, and converted them to strain. The resulting strain maps are shown in Figures 3c and 3d, respectively. Axial strain values relative to a $D_{gs}$ parameter of 2.87 GHz are shown. For non-axial strain, the minimum detectable resonance splitting is set by the magnetic field induced from the NV-site $^{14}$N nuclear spin\cite{Dolde2011}, allowing for absolute calibration of a zero value. The possible presence of multiple NV orientations could lead to a worst-case underestimation of strain through both the axial and non-axial parameters by a factor of two. Figure 3c indicates a maximal axial strain of $6\cdot10^{-4}$ at the grain boundary which relaxes towards the center of the grain, while Figure 3d indicates a maximum non-axial strain of $1.8\cdot10^{-4}$. These strains relax to their minimum value over a distance of 24 $\upmu$m from the grain boundary, setting a relevant length scale of the use of PCD in device design.

The sensitivity of this technique can be characterized by the 68\% confidence interval on the fitted resonance frequencies, which corresponds to the measurement standard deviation and depends primarily on the detected photon rate. Due to the non-uniform illumination across the field of view as well as background fluorescence near grain boundaries, the sensitivity varies spatially. The mean 68\% confidence interval on the two resonance frequencies for each 320 x 320 nm$^2$ pixel in the field of view is displayed in Figure 4. We achieve a median per-pixel ODMR spectral resolution of 245 kHz, resulting in an axial strain measurement precision of $2.7\cdot10^{-5}$ (nonaxial $1.2\cdot10^{-5}$) and a corresponding sensitivity of $3.21\cdot10^{-4}$ Hz$^{-1/2}$ (non-axial $1.5\cdot10^{-4}$ Hz$^{-1/2}$) for a 150 second measurement. To characterize a best-case sensitivity without the sample-specific background near the grain boundaries, we focus on the high-SNR regions corresponding to in-focus individual NVs at the center of the field of view. In these regions, we achieve 68\% confidence intervals of 79 kHz, corresponding to a measurement precision of $8.2\cdot10^{-6}$ (non-axial $3.8\cdot10^{-6}$) and sensitivity of $1.02\cdot10^{-4}$ Hz$^{-1/2}$ (non-axial $4.7 \cdot10^{-5}$ Hz$^{-1/2}$).

\begin{figure*}[h!]
	\includegraphics[width=17cm]{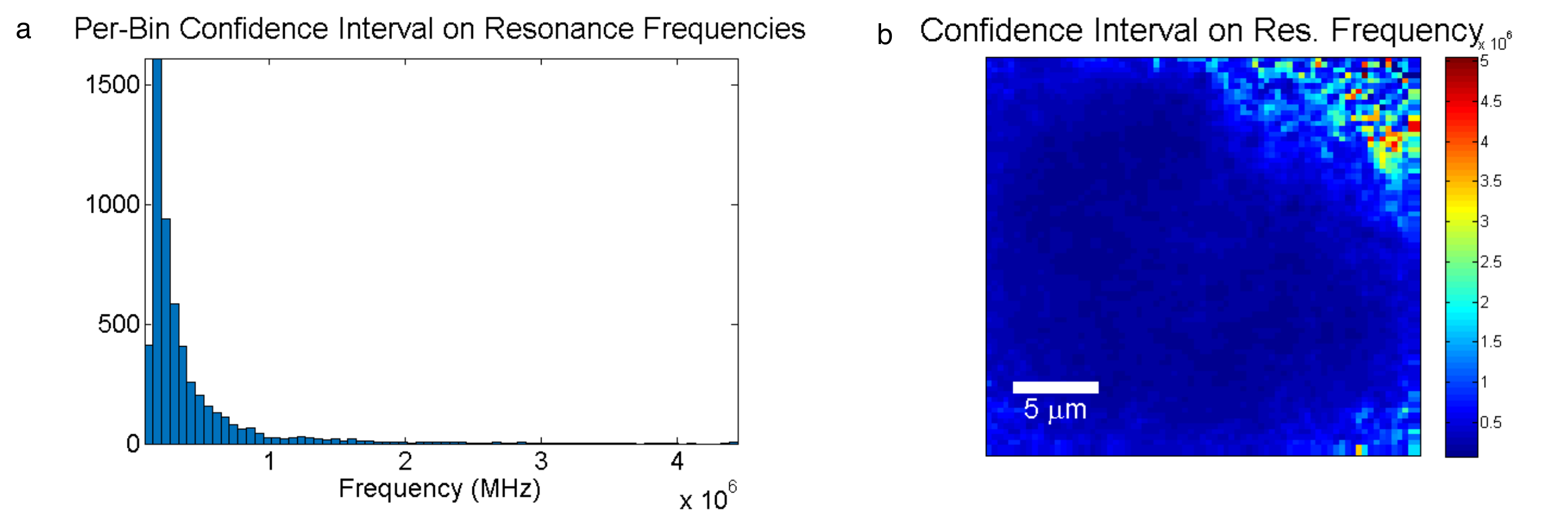}
	\caption{a)Histogram of the 68\% confidence interval on the fitted resonance frequencies for each
pixel of the strain image shown in the main text Figure 3. b) Spatial distribution of resonance fitting confidence intervals.
}
	\label{fig:spectrum}
\end{figure*}

Using this low magnetic field ODMR technique, we performed wide-field strain imaging in three dimensions across several crystal grains. These studies again reveal a strong strain gradient near grain boundaries. In the grain shown in Figure 3a-c, the D$_{gs}$ parameter corresponding to axial strain is lower near the boundary than in the center of the grain. This corresponds to tensile rather than compressive strain, while the reverse is true for the grain shown in Figure 3c,d. The non-axial strain at the center of the field of view is reduced with increased depth into the diamond, from a maximal value $ > 1\cdot10^{-4}$ near the surface to $ < 5\cdot10^{-5}$ at a depth of 3 $\upmu$m. As visible in the fluorescence profile (Figure 5a), the location and angle of the grain boundary varies with depth into the diamond, which is in turn reflected by rotation in the axial strain profile (Figure 5b,c). Here the strain is observed to relax within 10 $\upmu$m from the grain boundary. Further strain images, including extreme strain gradients and comparisons to reference single-crystal diamonds, are included in the Supplemental Material. These maps demonstrate the power of this technique for imaging crystal strain with high precision in three dimensions.

\begin{figure*}
	\includegraphics[width=17cm]{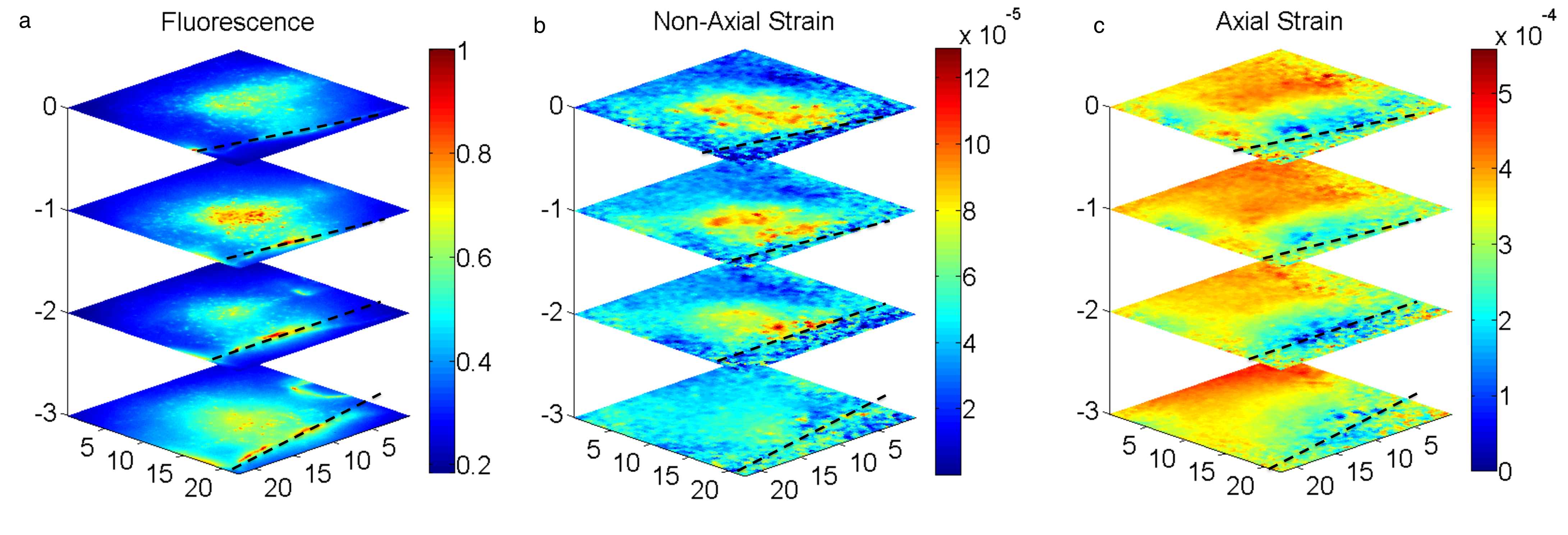}
	\caption{Three-dimensional strain imaging. a) Fluorescence b) Non-axial strain and c) Axial strain. The vertical images are offset by 1 $\upmu$m, with the top image on the surface of the diamond. The black line indicates the location of the grain boundary, and the spatial units are microns.}
	\label{fig:spectrum}
\end{figure*}

\section{Discussion}

Our studies reveal advantages and disadvantages of PCD for NV-based applications. The observation of preferential NV alignment within PCD grains, in combination with consistently high NV spin coherence times similar to single-crystal samples, can greatly improve the signal-to-noise of NV sensing applications (i.e a fourfold improvement in ODMR contrast)\cite{Pham2012a}; the large areas available for PCD diamond add to the usefulness in wide-field sensing applications, for example in biology\cite{Barry2000}. Strong strain gradients could also allow for sub-diffraction resolution of individual NVs using ODMR\cite{Chen2013}. In photonic devices, well-defined angular alignment of NVs can improve NV-dipole coupling to optical modes. PCD enables the production of wafer-scale photonic circuits in diamond, but roughness and scattering from grain boundaries have limited device performance compared to single-crystal diamond\cite{Rath2015a}. Large grains with low intra-grain roughness, such as those present in the sample characterized in this work, may reduce loss while still providing large-area substrates. Wide-field pre-characterization of the PCD could also allow for identification of and design compensation for grain boundaries.

Polycrystalline diamond could also serve as a natural environment for studying correlations between crystal strain and the spin and optical properties of embedded emitters. For example, high strain could break the orbital degeneracy of the SiV \cite{Muller2014a}, changing spin and orbital coherence properties, or strain gradients could be employed to tune relative NV sensitivity to electric and magnetic fields\cite{Jamonneau2015}. PCD provides a substrate that is ultra-pure and enables long spin coherence times, in contrast to other highly strained host materials such as nanodiamond. The strain in PCD can be very large; the observed NV ground state spin strain shifts of over 8 MHz in the axial direction are higher than those reported in cantilever-based experiments\cite{Teissier2014,Ovartchaiyapong2014} and would require pressures in the tens of GPa to produce externally. 
\section{Conclusion and Outlook}
The NV-based strain imaging technique introduced in this work reaches the optical diffraction limit and a high sensitivity of $10^{-5}$ Hz$^{-1/2}$ ($<$ 10 MPa), which outperforms more traditional strain imaging techniques such as Raman imaging\cite{Kato2012109} that are limited to absolute sensitivities $>$ 10 MPa in addition to being orders of magnitude slower \cite{Mermoux2004}. Through sectional imaging, this 3D imaging technique also offers an advantage over birefringence\cite{Pinto2009,Hoa2014} or x-ray topography\cite{Umezawa2011523} strain imaging methods, which image whole-sample and near-surface strains, respectively.  Although limited to diamond and other materials with optically accessible, strain-sensitive spin defects (e.g. silicon carbide\cite{Falk2014}), this technique has potential for precisely characterizing internal strains, such as those induced by geological formation\cite{Vlasov2013} or in strain-engineered devices, as well as for mapping externally applied strains. More advanced dynamical-decoupling sequences can increase axial strain sensitivity\cite{Brunner2013,Toyli2013}, which in turn could enable sub-diffraction strain imaging by resolving NVs in the spectral domain. While this work images individual NVs, higher-NV-density samples would increase sensitivity\cite{2008.NPhys.Taylor}, potentially enabling the imaging of few-site dislocations in single-crystal diamond or increasing the field of view. By imaging NVs in different independent orientations, this technique additionally could provide full vector reconstruction of the local strain. Since the NV is a truly nanoscale sensor, operable over a wide range of temperatures\cite{Toyli2012} and pressures\cite{Doherty2014}, this method can be used to perform ultra-high resolution strain mappings dynamically in previously unexplored regimes.
\\

\section{Supplemental Material}
See the supplemental material for a description extended discussions of preferential alignment and spin coherence times as well as additional strain imaging datasets.

\section{Acknowledgements}
The authors would like to thank Daniel Twitchen and Matthew Markham for helpful discussions, as well as C. Foy and H. Clevenson for their perspective on the manuscript.
This work was supported in part by the Air Force Office of Scientific Research (AFOSR) MURI (FA9550-14-1-0052), the AFOSR Presidential Early Career Award (supervised by Gernot Pomrenke), the Army Research office MURI biological transduction program, and the Office of Naval Research (N00014-13-1-0316).

\section{References}

\bibliography{library}

\clearpage

\title{Supplemental Material for Wide-Field Strain Imaging with Preferentially-Aligned Nitrogen-Vacancy Centers in Polycrystalline Diamond}

\section{Preferential Alignment under Varying Magnetic Fields}
One possible explanation for the appearance of less than four resonances in the NV ODMR spectrum under an external magnetic field is that fortuitous alignment of the field has resulted identical Zeeman shifts for different geometric classes of NV centers. To demonstrate that this is not the case, we vary the external field orientation relative to the sample and observe only a single resonance line under all magnetic field conditions, varying in frequency due to different magnetic field projection along the NV axis.

\begin{figure}[h!]
	\includegraphics[width=19cm]{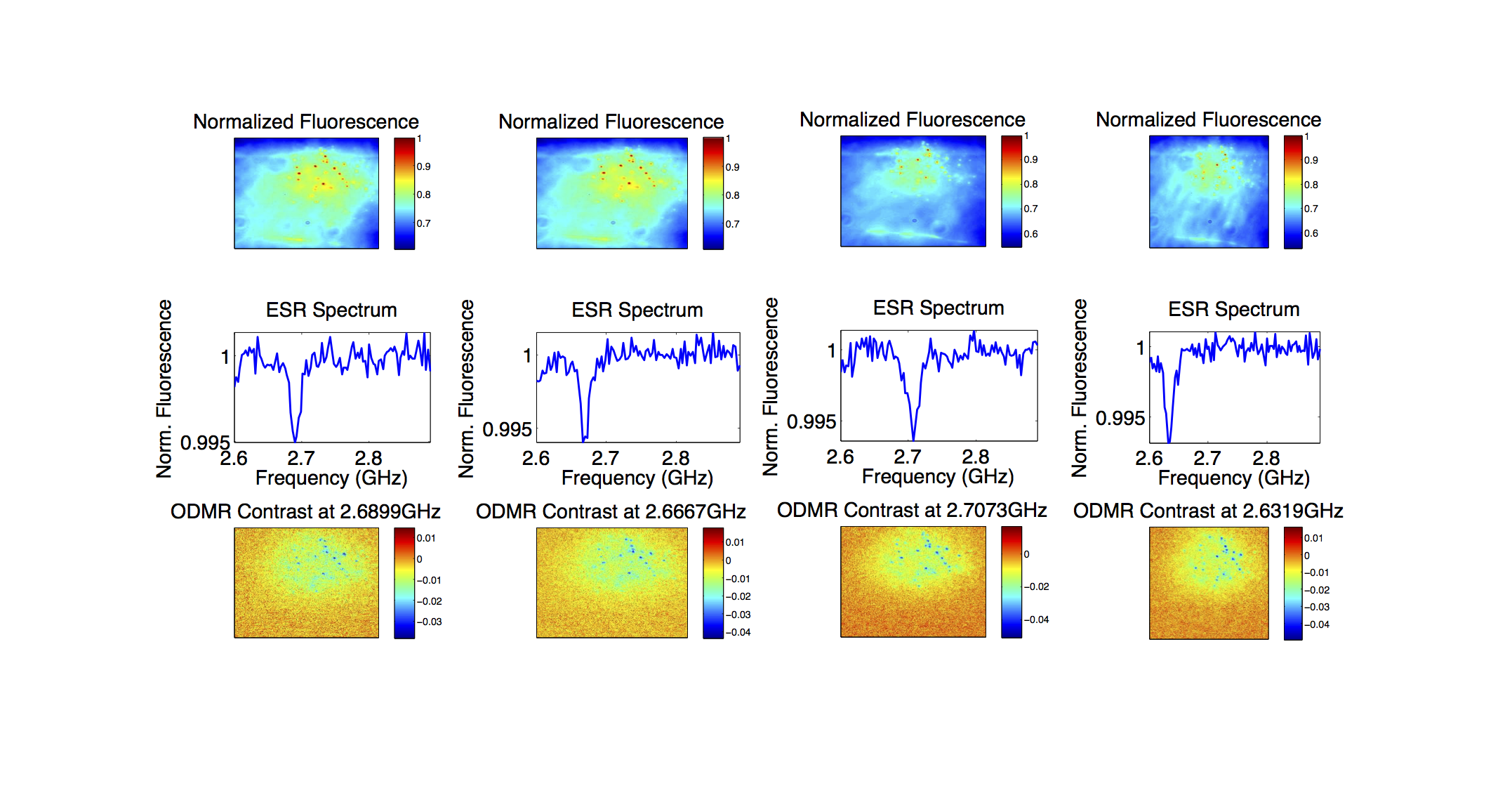}
	\caption{NV ODMR under different angular alignments of the external magnetic field (left to right). Top Row: Fluorescence. Middle Row: ODMR spectra averaged across the entire field of view. Bottom Row: ODMR contrast maps at the NV resonance frequency for each magnetic field orientation. While the area imaged is constant both in PL and contrast, only one resonance line is seen for all magnetic field angles.}
	\label{fig:spectrum}
\end{figure}

\section{NV Spin Coherence}

With the NV density observed to vary significantly across the PCD sample, a natural question is whether NV spin properties likewise change due to the apparent difference in paramagnetic defect density. To address this, we performed Hahn Echo measurements in regions with differing NV densities. Figure S2 shows the PL and spin data for each region represented in main text Figure 2f. The NV density was determined by counting the number of in-focus NV centers and dividing by the field of view, while the $T_2$ was computed through single-exponential fitting of revival decay envelope. We found that NV T$_2$ time is not strongly dependent on NV density, indicating that spin-bath correlation time does not significantly differ. In turn, this implies NV formation probability is dependent on crystal region.

\begin{figure*}[h!]
	\includegraphics[width=17cm]{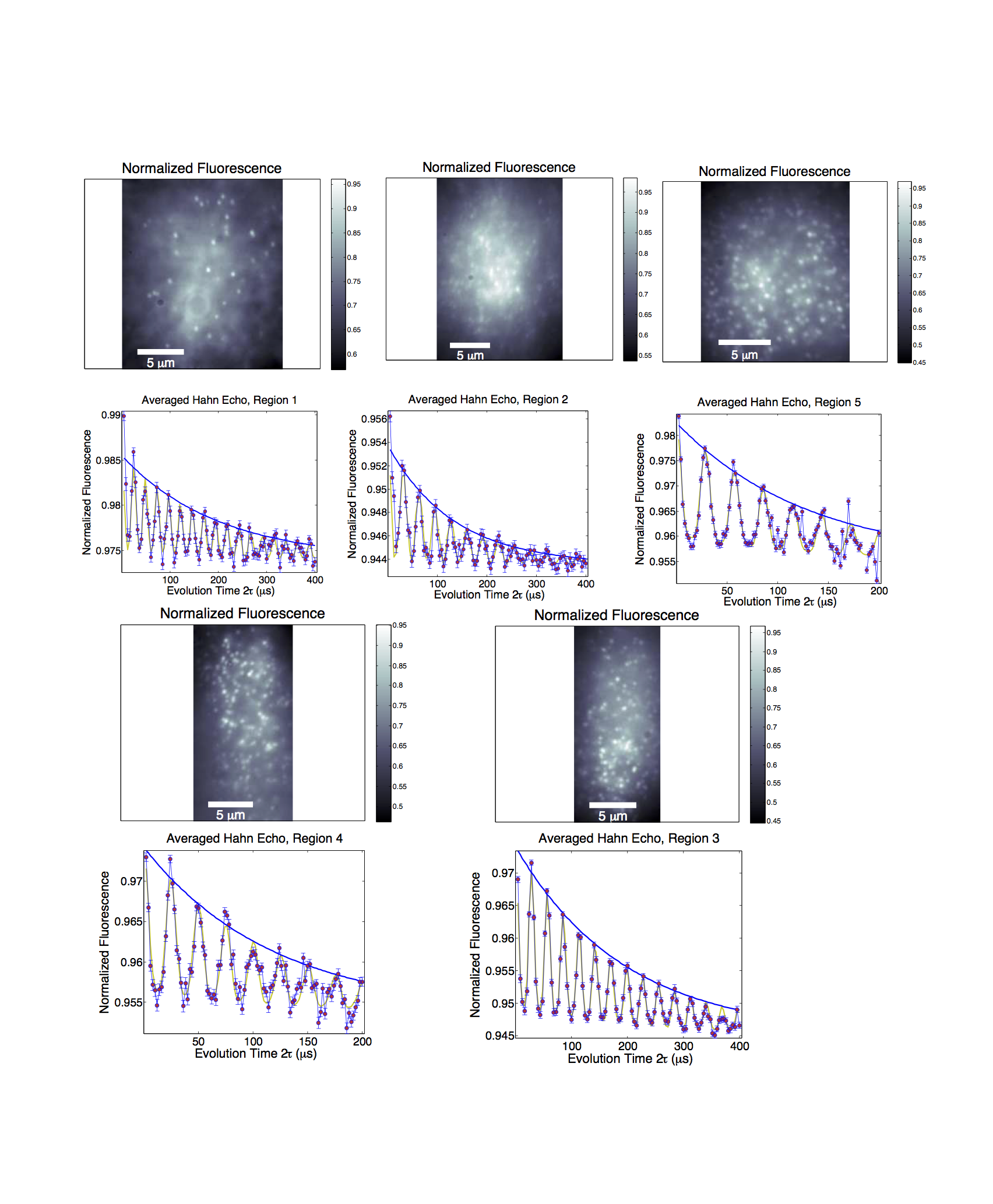}
	\caption{NV PL and Hahn echoes for 5 measured regions in PCD.}
	\label{fig:spectrum}
\end{figure*}

\begin{figure*}[h!]
	\includegraphics[width=17cm]{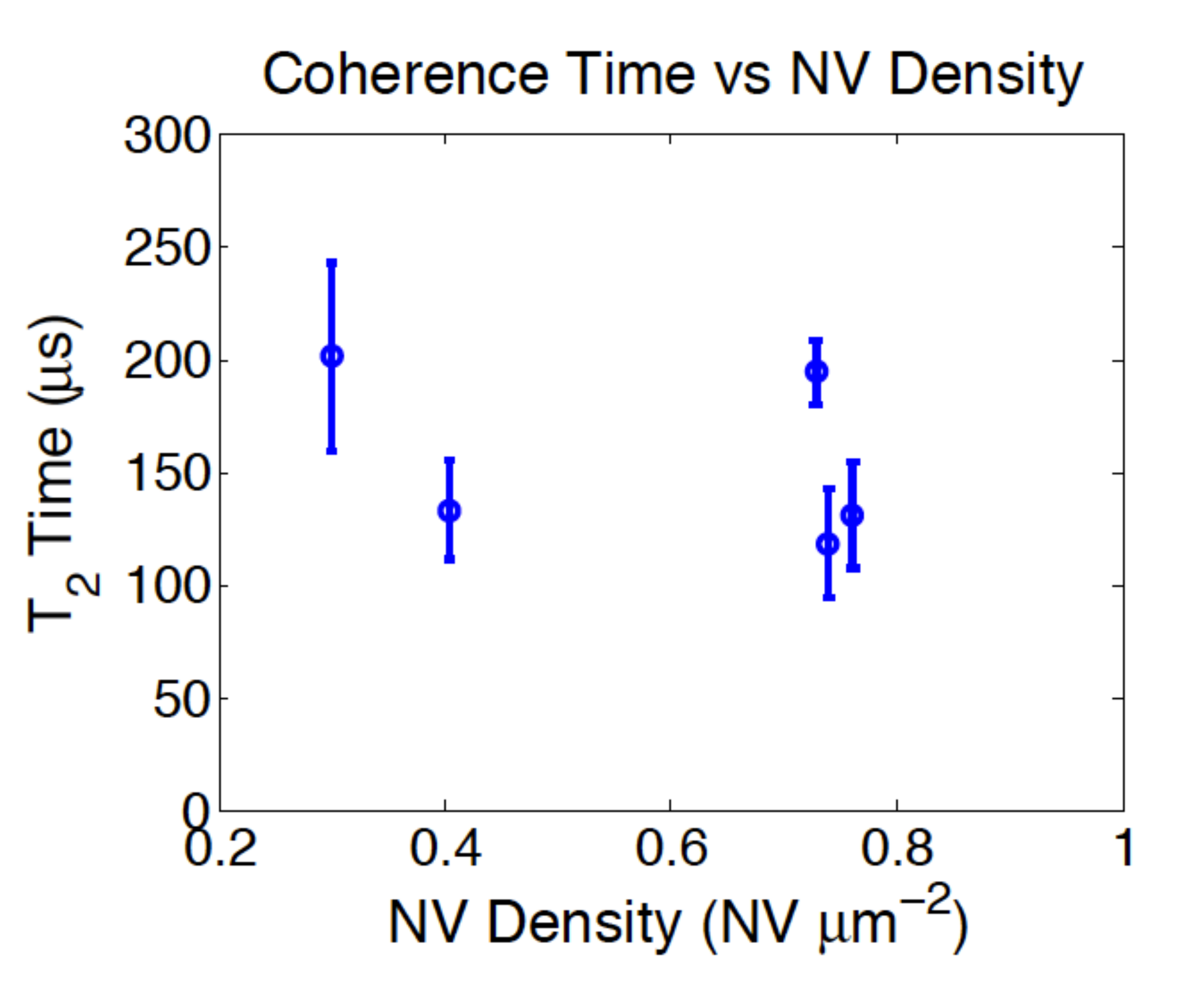}
	\caption{Dependence of coherence time on NV density}
	\label{fig:spectrum}
\end{figure*}

\section{Reference Measurements on Single-Crystal Diamond}

We use a single-crystal type IIa diamond with $\sim1$ ppm N (element6) as a reference sample. At zero magnetic field, we observe overlapping $^{14}$N hyperfine transitions from all four NV geometric classes. The inner resonances $m_i = 0$ are split by non-axial strain, while the outer resonances $m_i = \pm1$ are degenerate for $\vec{B} = 0$. The linewidth of the outer resonances, $\sim300$ kHz, therefore sets the minimum detectable magnetic field (or alternately, the maximum residual field), while the separation of the inner resonances corresponds to the average strain across the field of view of $1.9(5)\cdot10^{-5}$.

\begin{figure*}[h!]
	\includegraphics[width=17cm]{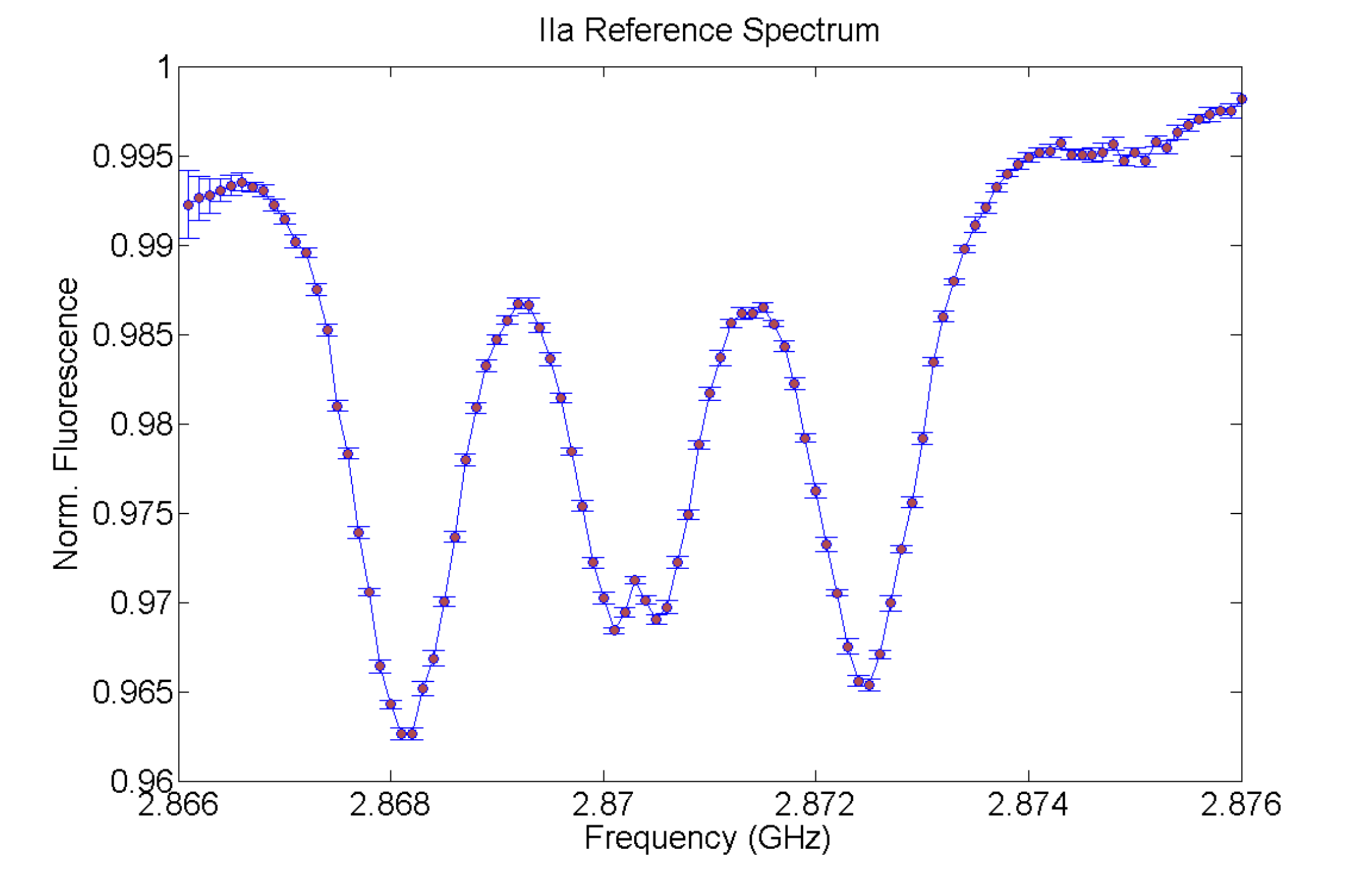}
	\caption{ODMR spectrum of an ensemble of NVs in a type IIa single-crystal reference diamond at zero external magnetic field}
	\label{fig:spectrum}
\end{figure*}

\section{High Strain Gradients}
We observe extremely high strain gradients of over one part in a thousand strain across 10 microns. This corresponds to pressures $ >$ 1 GPa. Remarkably, NV centers remained stable and displayed ODMR despite this wide change. Polycrystalline diamond could offer a testbed for the functionality of quantum devices in these extreme operational regimes. 
\begin{figure}[h!]
	\includegraphics[width=17cm]{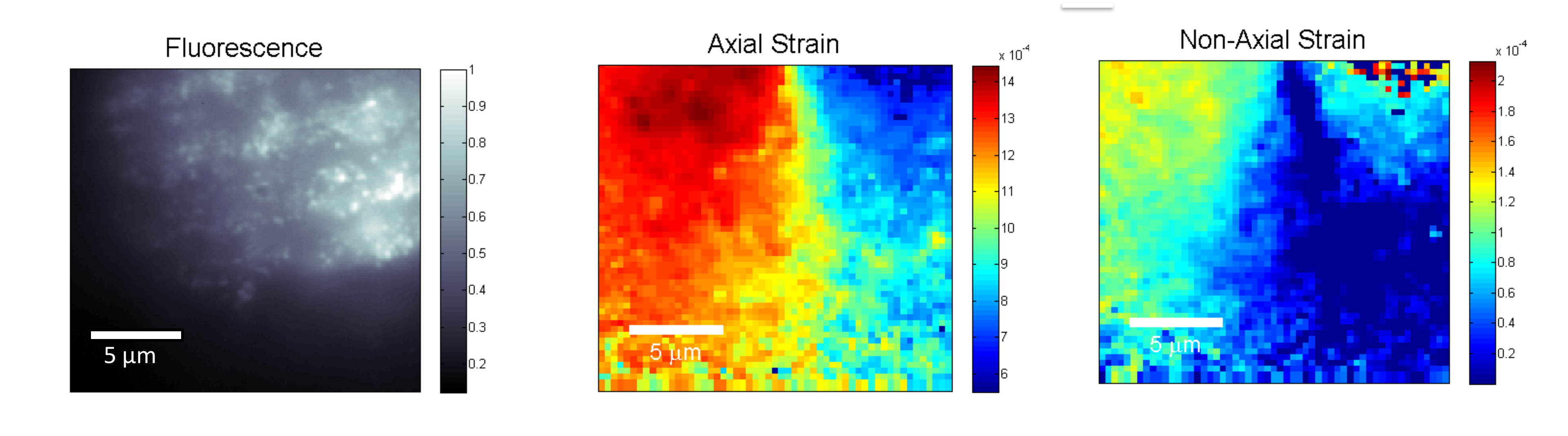}
	\caption{Highly strained PCD. a) NV PL. b) Axial and c) non-axial strain}
	\label{fig:spectrum}
\end{figure}

\section{Strain Measurement in the Presence of Multiple NV Orientations}

The description given in the main text is accurate in the single-NV case and generalizes directly to ensembles of NVs with identical orientation. In the case of multiple NV orientations, the resonance frequencies $\omega_{\pm}$ are different for each orientation as the projection of magnetic field and strain along the NV axis differs for each orientation. If the individual resonance lines for each orientation are resolvable the local field along and perpendicular to each NV direction can be reconstructed, in principle allowing for vector imaging. In the case that the resonance lines are not individually resolvable (i.e. the resonance linewidth is greater than the difference in the resonance frequencies between orientations), the observed resonance frequencies are the weighted average of the constituent resonance frequencies, where the weighting is given by the relative frequency of each orientation within the ensemble: \begin{equation}\hbar\omega_{\pm,obs} = \frac{1}{N}\Sigma_{i=1}^N F_i\omega_{\pm,i} =\frac{1}{N}\Sigma_{i=1}^N F_i (D_{gs}+\mathcal{E} _{z,i} \pm \sqrt{\mathcal{E} _{\perp,i}^2+(g\mu_bB_{z,i})^2})\end{equation}

We determine the axial strain by taking the mean of the two resonance frequencies, and the non-axial strain by the difference, in the limit of no magnetic field. Because there is a linear relation between resonance frequency and strain in this limit, the observed strain is the weighted average of the true axial and non-axial strains for each orientation.

\begin{equation} \mathcal{E} _{z,obs} = \frac{\hbar}{2}(\omega_{+,obs}+\omega_{-,obs}) = \frac{1}{N}\Sigma_{i=1}^N F_i (D_{gs}+\mathcal{E} _{z,i}) \end{equation}
\begin{equation} \mathcal{E} _{\perp,obs} = \frac{\hbar}{2}(\omega_{+,obs}-\omega_{-,obs}) = \frac{1}{N}\Sigma_{i=1}^N F_i \mathcal{E} _{\perp,i} \end{equation}

In this limit, therefore, we measure the strain components relative to the mean NV axial and non-axial moments.

\end{document}